\documentclass[reprint,amsmath,amssymb,aps,a4paper]{revtex4-1}
\usepackage{graphicx}
\usepackage[utf8]{inputenc}
\usepackage{braket}
\usepackage{comment}
\usepackage{float}
\usepackage{siunitx}
\usepackage[colorlinks=true, linkcolor=blue, urlcolor=blue, citecolor=blue]{hyperref}

\begin{document}

\title{Generation of Schrödinger cat states\\ through photon-assisted Landau-Zener-St\"{u}ckelberg interferometry}

\author{Jonas Lidal}
\author{Jeroen Danon}
\affiliation{Center for Quantum Spintronics, Department of Physics, Norwegian University of Science and Technology, NO-7491 Trondheim, Norway}

\date{\today}

\begin{abstract}
Schrödinger cat states are useful for many applications, ranging from quantum information processing to high-precision measurements.
In this paper we propose a conceptually new method for creating such cat states, based on photon-assisted Landau-Zener-Stückelberg interferometry in a hybrid system consisting of a qubit coupled to a photon cavity.
We show that by initializing the qubit in one of its basis states, performing three consecutive sweeps of the qubit energy splitting across the 1-photon resonance, and finally projecting the qubit to the same basis state, the parity of the photon field can be purified to very high degree; when the initial photon state is a coherent state, the final state will then be very close to a Schrödinger cat state.
We present numerical simulations that confirm that our protocol could work with high fidelity ($\sim 0.99$) for coherent states of reasonable size ($|\alpha|^2 \sim 10$).
Furthermore, we suggest that our protocol can also be used to transfer quantum information between the qubit and a superposition of orthogonal cat states in the cavity.
\end{abstract}

\maketitle

\section{\label{sec:intro}Introduction}
A coherent state is a quantum state of the harmonic oscillator that most closely resembles a classical state, in the sense that it has minimal and equal uncertainty in its two conjugate variables, the expectation values of which follow the classical equations of motion.
More explicitly, coherent states are the eigenstates of the oscillator's bosonic annihilation operator, $a \ket{\alpha} = \alpha \ket{\alpha}$, where $\alpha$ is a complex number characterizing the amplitude and phase of the oscillations associated with $\ket{\alpha}$: $|\alpha|^2$ gives the expectation value of the number of excitations in the oscillator.

A superposition of two coherent states, e.g.,
\begin{equation}\label{SCS}
\ket{\Psi_\pm(\alpha)} = \frac{1}{\cal N} \left( \ket{\alpha} \pm \ket{-\alpha} \right),
\end{equation}
is, in a way, thus analogous to the cat in Schrödinger's famous thought experiment \cite{Schrodinger1935}, as it presents a quantum superposition of two different (quasi-)classical states.
These superpositions are therefore commonly known as Schrödinger cat states and are interesting for a number of reasons.
Firstly, since their behavior is on the border between quantum and classical, they provide a perfect playground for studying decoherence and the quantum-to-classical transition \cite{Brune1996}, which is of fundamental interest.
Further, it has been shown that Schrödinger cat states can be used as a resource for quantum computation \cite{Ralph2003,Lund2008} and quantum error correction \cite{Cochrane1999,Gottesman2001,Bergmann2016,Ofek2016}, quantum teleportation \cite{VanEnk2001,Jeong2001}, and also high-precision measurements \cite{Munro2002,Ralph2002a,Gilchrist2004,Joo2011}.
For these reasons, reliable generation and manipulation of such cat states has been the focus of a substantial amount of work in the past few decades, both theoretical and experimental.

Most of the cat-based quantum technologies mentioned above, such as high-precision metrology and reliable quantum computation, require the use of coherent states of the freely propagating photon field.
Furthermore, these applications work best when the overlap between the coherent states constituting the cat state is small.
For the state (\ref{SCS}) this overlap is $|\braket{\alpha|-\alpha}| = e^{-2|\alpha|^2}$, and in that case it has been estimated that $|\alpha|>1.2$ is required for fault-tolerant quantum computing \cite{Lund2008}.

Over the years, many ways have been put forward how to produce freely propagating cat states, several of which have successfully been implemented.
Yurke and Stoler originally proposed sending a coherent photon state through a strongly non-linear (Kerr) medium to generate a cat state~\cite{Yurke1986a}, but all commonly available media are too weakly non-linear to achieve the required degree of dispersion over reasonable distances.
Other proposed methods, some of which have been successfully implemented, include
performing conditional measurements on the squeezed vacuum~\cite{Dakna1997,Ourjoumtsev2006,Neergaard-Nielsen2006,Takahashi2008,Ourjoumtsev2009,Gerrits2010a},
mixing a coherent state with a squeezed single-photon beam~\cite{Lund2004},
homodyne detection on a 50/50 split $n$-photon Fock state~\cite{Ourjoumtsev2007},
and reflecting coherent light pulses from an atom-cavity system \cite{Hacker2019}.
The drawback of these methods is that they become less successful for increasing amplitude $|\alpha|$ of the cat state, the highest amplitudes reached being $|\alpha| \sim 1.5$. 

Achieving higher amplitudes is possible via a few different methods.
One idea is to combine pairs of small-amplitude cat states into one state with a larger $|\alpha|$ in a process known as ``breeding'' \cite{Lund2004,Sychev2017}.
Another route is to turn to cavity-QED, trapping the photons in cavities where they strongly interact with atoms that are shot through the cavity; in that way, the state of the photon field can be manipulated into a cat state~\cite{Brune1992,Brune1996,Deleglise2008a}.
Advances in qubit technology allowed for replacing the atoms with (superconducting) qubits acting as artificial atoms that are coupled to the cavity mode, which provides an extra level of control over the light-matter interaction.
Such hybrid systems can be used to coherently transfer quantum information from a qubit to a superposition of cat states~\cite{Leghtas2013} and they allowed for the creation of cat states with amplitudes up to $|\alpha| \sim 10$~\cite{Vlastakis2013}.

Such cavity-based cat states were used for studying the quantum-to-classical transition~\cite{Brune1996}, but they could also provide a platform for fault-tolerant quantum information processing~\cite{Gottesman2001,Leghtas2013}.
Furthermore, for applications where a freely propagating cat state is needed it is possible to ``release'' a non-classical photon state from a cavity, which has been demonstrated for single-photon~\cite{Yoshikawa2013} as well as multi-photon states~\cite{Pfaff2017}.

In this paper, we propose a conceptually new method of generating Schrödinger cat states, based on photon-assisted Landau-Zener-St\"{u}ckelberg interferometry in a hybrid system consisting of a qubit coupled to a photon cavity. Landau-Zener-Stückelberg interferometry has been used before to create entangled states in multi-partite few-level systems \cite{sun2010,sun2011}, but here we explore the possibility to use it to entangle the \emph{photon field} by manipulating the level structure of the qubit. We show that repeatedly sweeping the level splitting of the qubit through the 1-photon resonance can lead to interference effects which, depending on the details of the level crossing, can selectively amplify and attenuate specific $n$-photon components in the wave function of the cavity field.
Using this principle, we demonstrate how an initial coherent photon state in the cavity can be transformed with high fidelity to a so-called even or odd cat state $\ket{\Psi_\pm(\alpha)}$ [see Eq.~(\ref{SCS})] by means of three consecutive level crossings.
We also speculate that the procedure can be used to transfer quantum information from a qubit state to the photon cavity using an odd cat state and an even state as a basis.
We further present numerical simulations of the time-evolution of the proposed system which confirm the successful creation of cat states with fidelities up to $\sim 0.99$ for $|\alpha|^2 \sim 10$.

The rest of this paper is organized as follows.
In Sec.~\ref{sec:LZSI} we review the basics of Landau-Zener-Stückelberg interferometry, using the example of two levels that are swept through each other multiple times.
In Sec.~\ref{sec:model} we then introduce the system and model Hamiltonian we consider, and we outline the basic working of our cat-state generating protocol; first we do this using the most intuitive picture possible, connecting directly to the example system presented in Sec.~\ref{sec:LZSI}, and then we discuss the main simplifications we made in that picture and estimate the deviations from the ideal situation in a more realistic picture.
We end the Section suggesting how the same protocol could be used to coherently transfer quantum information from the qubit to a superposition of cat states in the cavity.
In Sec.~\ref{sub:numerics} we present our numerical simulations, which confirm the working of protocol.
Finally, in Sec.~\ref{sec:realization} we discuss a few candidate systems that could be used to implement our idea and in Sec.~\ref{sec:conc} we present our conclusions.

\section{\label{sec:LZSI}Landau-Zener-Stückelberg interferometry}

The dynamics of a time-dependent level crossing in a two-level system is a well studied problem in quantum mechanics and can be described by the Hamiltonian
\begin{equation}\label{qubitHam}
	H_{\rm qub} = \frac{\Delta(t)}{2} \sigma_z + \delta\, \sigma_x,
\end{equation}
written in the diabatic basis $\{\ket{1},\ket{0}\}$ and using the Pauli matrices $\sigma_{x,z}$.
Assuming linear driving of the level splitting, $\Delta(t) = vt$, where $t$ is time and $v$ the sweep speed, the system will pass a region around $t=0$ where the coupling term $\delta$ mixes the two components of the wave function.
In the limit of an infinite linear sweep of the energy splitting, from $t = -\infty$ to $t=\infty$, the probability of a diabatic transition (i.e., the probability for the system to remain in its initial state after the crossing) is given by the famous Landau-Zener formula,
\begin{equation}\label{LZprob}
	P_{\rm LZ}=e^{-2\pi\frac{\delta^2}{\hbar v}},
\end{equation}

This analytic result is valid for an infinite sweep through a single level crossing, and thus needs to be adapted to describe the situation of multiple consecutive level crossings.
Assuming that the crossings are far enough apart in time, it is reasonable to assume that all crossings can be treated separately, leading to the adiabatic-impulse model~\cite{Damski2006}.
The general idea is to treat the system as if it evolves adiabatically everywhere except in regions close to the level crossings and the non-adiabatic evolution at all crossings is assumed to be instantaneous.
This approximation is considered good if the crossings are locally linear in time and well separated \cite{Shevchenko2010}.

The time-evolution operator can then be written as a series of adiabatic evolution operators separated by non adiabatic transfer operators. The adiabatic evolution operator, in the \emph{adiabatic} basis, is
\begin{equation}
	U(t_2,t_1) =\begin{pmatrix}
					e^{-i\theta_+(t_2,t_1)} && 0\\
					0 && e^{-i\theta_-(t_2,t_1)}
				\end{pmatrix},\label{eq:u}
\end{equation}
where $\theta_\pm(t_2,t_1) = \int_{t_1}^{t_2} dt\, E_\pm(t)$ in terms of the instantaneous eigenenergies $E_\pm(t)$.
The non-adiabatic evolution at the level crossing reads in the same basis as~\cite{Shevchenko2010}
\begin{equation}
	N= \begin{pmatrix}
		\sqrt{1-P_{\rm LZ}}e^{-i{\phi}_{\rm S}} && -\sqrt{P_{\rm LZ}}\\
		\sqrt{P_{\rm LZ}}&&\sqrt{1-P_{\rm LZ}}e^{i{\phi}_{\rm S}}
		\end{pmatrix},
\end{equation}
where we see that, apart from the (square root) of the Landau-Zener probabilities, all amplitudes pick up a different phase, where $\phi_S$ in the diagonal elements is~\cite{Shevchenko2010},
\begin{equation}\label{stokesPhase}
	\phi_{\rm S} = -\frac{\pi}{4} + \frac{\delta^2}{\hbar v}\left[ \ln \left( \frac{\delta^2}{\hbar v}\right) -1\right] + \arg\,\Gamma\left(1-i\frac{ \delta^2}{\hbar v}\right),
\end{equation}
with $\Gamma(z)$ being the gamma function.

\begin{figure}[t]
	\begin{center}\includegraphics[width=0.42\textwidth]{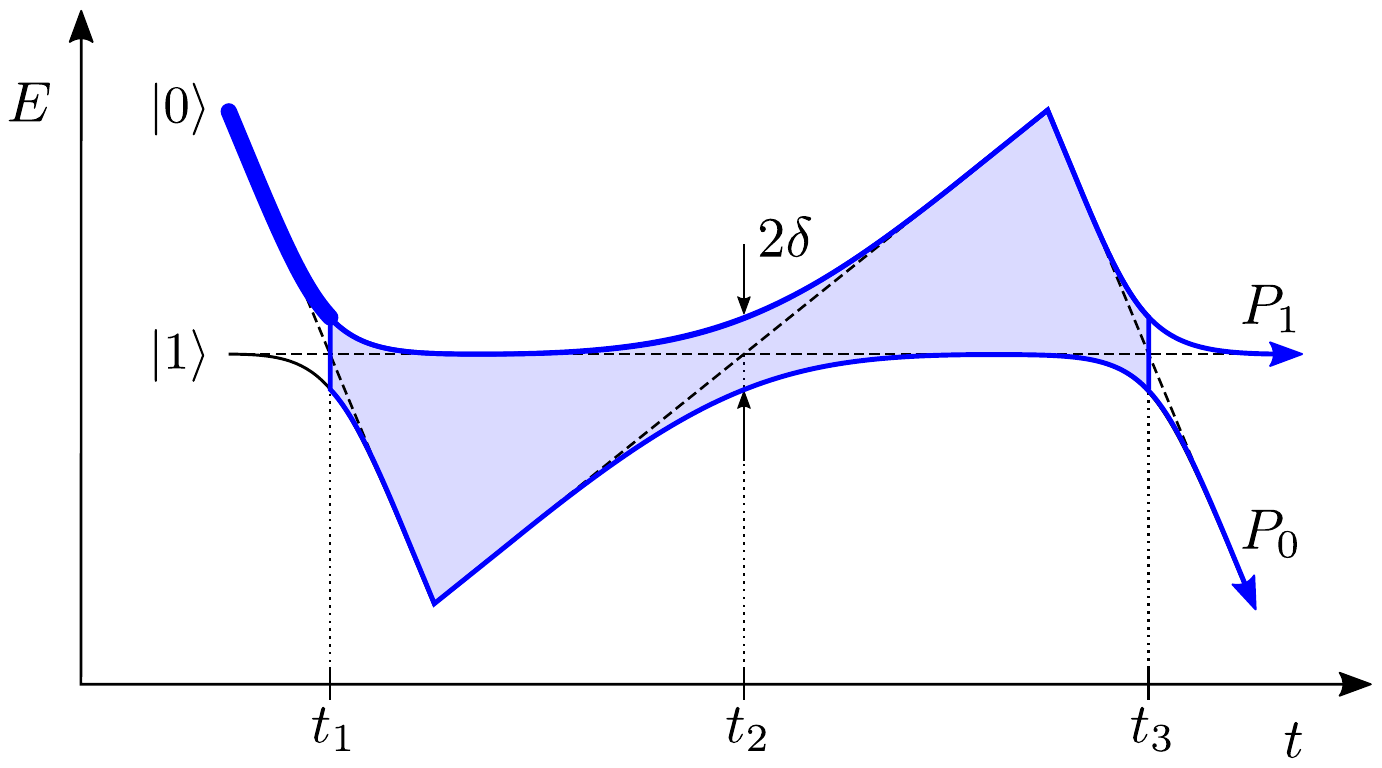}\end{center}
	\caption{Time-dependent spectrum of the example sweep protocol discussed in Sec.~\ref{sec:LZSI}.
		The diabatic energies of $\ket{0}$ and $\ket{1}$ are shown as dashed lines, while the adiabatic (instantaneous) eigenenergies are shown as full lines.
		Two interfering paths are colored in blue and the dynamic phase difference between them is shown as a blue shaded area.}
	\label{fig:LZSIexample}
\end{figure}

As an example we consider initializing the system in the state $\ket{0}$ and driving the level splitting through three consecutive crossings, as shown in Fig.~\ref{fig:LZSIexample}.
The first and last have identical sweep speed $v$, while the middle crossing is so slow that it is adiabatic.
In this limit, the matrix $N$ at the middle crossing becomes $i\sigma_z$ and the final state in the adiabatic-impulse model is thus given by
\begin{equation}\label{exampleEvo}
	\ket{\psi}_f = iU(t_f,t_3)N\sigma_zU(t_3,t_1)NU(t_1,t_i) \ket{0},
\end{equation}
where $t_i<t_1$ is the initial time and $t_f>t_3$ the final time.
For the case where $P_{\rm LZ}=\frac{1}{2}$ for the first and last crossing we can write explicitly in the adiabatic basis
\begin{align}
\ket{\psi}_f = {} & {} e^{-i[\theta_+(t_1,t_i)+\frac{1}{2}\theta_+(t_3,t_1)+\frac{1}{2}\theta_-(t_3,t_1)]} \nonumber\\
{} & {} \times \begin{pmatrix}
	i e^{-i[\theta_+(t_f,t_3)+\varphi]}\cos(\frac{1}{2}\phi_d+\varphi) \\
	e^{-i\theta_-(t_f,t_3)}\sin(\frac{1}{2}\phi_d+\varphi)
	\end{pmatrix},\label{eq:finalstate}
\end{align}
where $\phi_{d} = \theta_+(t_3,t_1) - \theta_-(t_3,t_1)$ is the phase difference built up during the adiabatic evolution from $t_1$ to $t_3$ and $\varphi \approx -1.08$ is the phase $\phi_{\rm S}$ that corresponds to the value of $\delta^2/\hbar v$ that yields the probability $P_{\rm LZ}=\frac{1}{2}$.
The final occupation probabilities for the two states then follow straightforwardly as
\begin{equation}
P_{1,0} = \frac{1}{2} \pm \frac{1}{2} \cos{(\phi_{d} + 2\varphi)}.\label{eq:p10simple}
\end{equation}

We thus see how this sweep protocol indeed leads to interference effects that depend on the difference in phases acquired along the two possible paths in time (blue lines in Fig.~\ref{fig:LZSIexample}).
This phase difference has two contributions: (i) the phase $\varphi$, given by (\ref{stokesPhase}), caused by the first and the last level crossing, and (ii) the dynamical contribution $\phi_{d}$ picked up during the adiabatic evolution, which corresponds to the blue shaded area in Fig.~\ref{fig:LZSIexample}.
As we can see, the total phase and hence the ``return'' probability is periodically dependent on this area, which we can easily control by tuning the sweep speed or the coupling strength $\delta$ at the second crossing.

\section{\label{sec:model}Proposal}

The goal is to create even or odd cat states $\ket{\Psi_\pm(\alpha)} \propto \ket{\alpha} \pm \ket{-\alpha}$ in the photon field, where the coherent states read explicitly as $\ket{\alpha} = e^{-|\alpha |^{2}/2} \sum _{n=0}^{\infty}(\alpha ^{n} /{\sqrt {n!}})\ket{n}$ in terms of the photon number basis states $\ket{n}$.
We see that, due to the factor $\alpha^n$ in the photon number coefficients, an even(odd) cat state only contains even(odd) photon number states, the occupation probabilities of which still have the same Poissonian ``envelope'' as the coherent states they are constituted of.
In short, our cat state generation protocol, which we will explain in detail below, amounts to removing all odd or even components from a coherent photon state, without altering the Poissonian envelope structure of the state too much.

Let us now turn to the hybrid qubit-cavity system with which we would like to perform this protocol.
We assume a simplest situation, where a single qubit is coupled to a single mode of the cavity photon field, and we describe the system with the Hamiltonian
\begin{equation}\label{transHamiltonian}
H = \frac{\Delta(t)}{2} \sigma_z + \hbar \omega a^\dagger a + \hbar A(a + a^\dagger)\sigma_{x},
\end{equation}
where $\Delta(t)$ is the time-dependent qubit splitting, $\omega$ the frequency of the cavity mode, and $A$ is the coupling strength between the qubit and the photon field. 
We note that we assumed ``transverse'' qubit-cavity coupling, i.e., the field in the cavity couples to the $\sigma_x$ operator in the qubit subspace; comparing with the toy model used in Sec.~\ref{sec:LZSI} we see that $\hbar A(a + a^\dagger)$ now takes the place of $\delta$.

\begin{figure}[t]
	\centering
	\includegraphics[width = 0.45\textwidth]{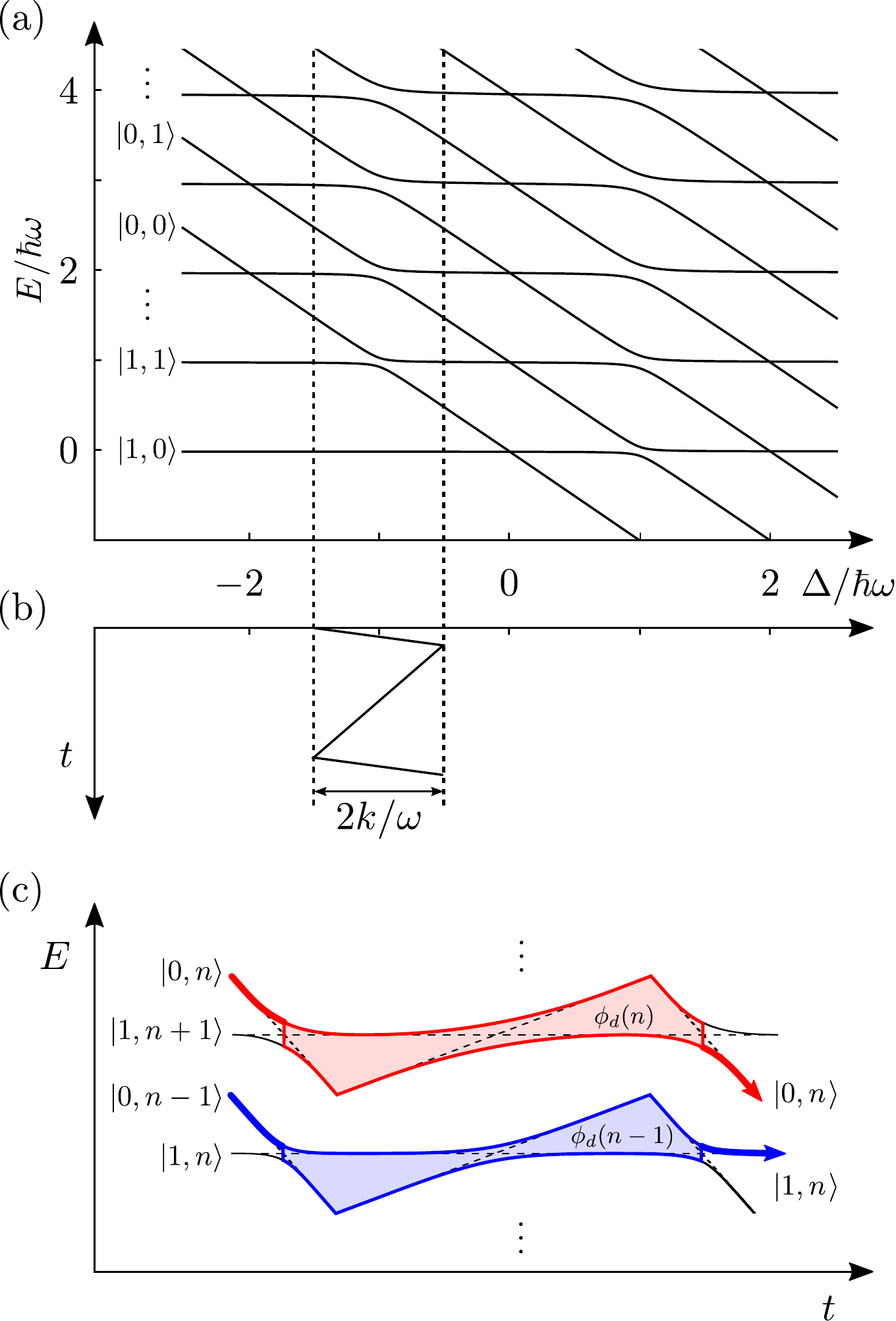}
	\caption{(a) Schematic plot of the energy spectrum of the Hamiltonian (\ref{transHamiltonian}) as a function of $\Delta$.
		Dressed levels $\ket 1$ and $\ket{0}$ anticross whenever the accompanying number of photons differs by 1.
		(b) Sketch of the proposed sweep protocol $\Delta(t)$.
		(c) Zoom in on neighboring pair of coupled levels that are swept through each other.
		The $n$-dependent magnitude of the relative dynamical phases picked up during the sweep is indicated by the shaded red and blue areas.}
	\label{fig:spectrum}
\end{figure}

In Fig.~\ref{fig:spectrum}(a) we sketch the spectrum of $H$ as a function of $\Delta$, where the labels $\ket{i,n}$ indicate the (approximate) basis states $\ket{i}_{\text{qub}} \otimes \ket{n}_{\text{ph}}$.
All levels cross, except when the two photon numbers involved differ by 1, in which case the magnitude of the anticrossing is determined by the matrix element
\begin{equation}
\braket{1, n|H|0,n\pm 1}
= \hbar A\sqrt{n+\tfrac{1}{2}\pm\tfrac{1}{2}},
\end{equation}
and thus depends on the actual photon numbers involved~\cite{Note1}.
The key to our proposal is to use this $n$-dependence of the size of the anticrossings.
If we drive the qubit in a zig-zag pattern around the 1-photon resonance, as sketched in Fig.~\ref{fig:spectrum}(b), then we create a ``ladder'' of time-dependent level crossings between levels $\ket{0,n}$ and $\ket{1,n+1}$, see Fig.~\ref{fig:spectrum}(c).
Each pair of coupled levels thus undergoes a sweep pattern that is similar to the one discussed in Sec.~\ref{sec:LZSI} and, assuming that we again make sure that the second level crossing is adiabatic, the final state is then given by Eq.~(\ref{eq:finalstate}), where the phase difference $\phi_d$ now depends on $n$, as can be seen from the difference in size between the red and blue shaded regions in Fig.~\ref{fig:spectrum}(c). We note that isolating the dynamics of pairs of levels in such a way is in fact equivalent to applying the usual rotating wave approximation to the Hamiltonian \eqref{transHamiltonian}.

The idea is then to initialize in $\ket{0}\otimes\ket{\alpha}$, i.e., the cavity in a coherent state and the qubit in one of its basis states, and perform the sweep protocol sketched in Fig.~\ref{fig:spectrum}, aiming for the characteristics (i) $P_{\rm LZ} = \tfrac{1}{2}$ for all first and last crossings, (ii) all middle crossings are adiabatic, and (iii) $\phi_d(n+1) - \phi_d(n) = \pi$ as closely as possible for all relevant $n$.
Of course, in reality the $n$-dependence of the dynamical phase difference is not linear over larger ranges of $n$ and $P_{\rm LZ}$ is also $n$-dependent.
Below we will investigate how well the desired characteristics can be satisfied at the same time, but let us for now assume the ideal (hypothetical) situation where $P_{\rm LZ} = \tfrac{1}{2}$ and $\phi_d(n) = \phi_d(0) + n\pi$.

The phase $\phi_d(0)$ can be tuned by changing parts of the time-dependent level structure that are the same for all $n$, such as the minimum and maximum value of $\Delta(t)$ or the coupling strength $A$.
Tuning $\phi_d(0)=-2\varphi$ is thus in principle possible, independently from trying to approach $\phi_d(n+1) - \phi_d(n) = \pi$.
Then we see from Eq.~(\ref{eq:finalstate}) that the initial component $\ket{0,n}$ will evolve as
\begin{align}
	\ket{0,n} \to e^{-in\omega T}\Big [ {} & {} \sin\left(\frac{n}{2}\pi\right) \ket{0,n} \nonumber\\
	{} & {}  + ie^{-i\varphi}\cos\left(\frac{n}{2}\pi\right)\ket{1,n+1} \Big],\label{eq:final0n}
\end{align}
where we introduced the total sweep time $T=t_f-t_i$, setting $t_f = t_3$ and $t_i = t_1$ for simplicity~\cite{Note2}.
We see that all components $\ket{0,n}$ evolve (up to a phase factor) into $\ket{0,n}$ for odd $n$ and into $\ket{1,n+1}$ for even $n$, thereby yielding zero weight in all even-$n$ components of the final state of the photon field~\cite{Note3}.

We thus end up with a photon field with a perfect odd parity.
Since the weight of each even-$n$ component that was removed was transferred to a neighboring $n$, one could expect that the envelope of the resulting cavity state is still relatively close to that of the coherent state, thus yielding (almost) a cat state.
Even better, however, would be to end the protocol with a selective measurement of the qubit state:
only accepting the outcome $\ket{0}$ will project the photon field to the state
\begin{align}
	\frac{-ie^{-\frac{1}{2}|\beta |^{2}}}{\cal N} {} & {}  \sum _{n=0}^{\infty}\left[ \frac{\beta^{n}}{\sqrt {n!}} -  \frac{(-\beta)^{n}}{\sqrt {n!}} \right] \ket{n} \nonumber\\
	{} & {} = -i \ket{\Psi_-(\beta)},\label{eq:finalcat}
\end{align}
with $\beta = ie^{-i\omega T}\alpha$, which is a perfect odd cat state.

Let us now investigate how closely our idealized assumptions about $P_{\rm LZ}$ and the dynamical phase differences can actually be met by a system described by the Hamiltonian \eqref{transHamiltonian}.
First of all, we write the full final state for the pair of levels $\{\ket{0,n}, \ket{1,n+1}\}$ after the sweep protocol, now allowing for deviations from our assumptions.
From Eqs.~(\ref{eq:u})--(\ref{exampleEvo}) we find the explicit expression
\begin{align}
	\ket{0,n} \to {} & {} e^{-in\omega T}\Big \{ 2 \sin\left(\tfrac{1}{2}\phi_d + \phi_{\rm S} \right)\sqrt{P_{\rm LZ}(1-P_{\rm LZ})} \ket{0,n} \nonumber\\
	 {} & {} + e^{-i\phi_{\rm S}} \Big[ (1-2P_{\rm LZ})  \sin\left(\tfrac{1}{2}\phi_d + \phi_{\rm S} \right) \nonumber\\
	{} & {} \quad +i  \cos\left(\tfrac{1}{2}\phi_d + \phi_{\rm S} \right) \Big] \ket{1,n+1}\Big\},\label{eq:final0nfull}
\end{align}
where now $\phi_d$, $\phi_{\rm S}$, as well as $P_{\rm LZ}$ are $n$-dependent.
As is well known, the relative width of the photon number distribution of a coherent state decreases for increasing $|\alpha|$, suggesting that this $n$-dependence might become less important for large $|\alpha|$.

We first focus on the $n$-dependence of the Landau-Zener probability for the first and last crossing, which simply reads as
\begin{align}\label{eq:PLZCond}
	P_{\rm LZ}(n) = \exp \left( -\frac{2\pi \hbar A^2[n+1] }{v_f} \right),
\end{align}
where $v_f$ is the ``fast'' sweep speed at those two crossings.
For any particular $\alpha$ we can tune this speed such that $P_{\rm LZ}(|\alpha|^2) = \frac{1}{2}$.
We then estimate the decrease in fidelity of creating the desired state due to the $n$-dependence of $P_{\rm LZ}$ by calculating the modulo square of the weight of the final component $\ket{0,n}$ at $n_\pm = |\alpha|^2 \pm |\alpha|$, using the fact that the photon distribution will have a Poissonian envelope with both mean and variance equal to $|\alpha|^2$.
We then find that for $|\alpha|^2 \gtrsim 1$
\begin{align}
	4 P_{\rm LZ}(n_\pm) [ 1-P_{\rm LZ}(n_\pm)] \approx 1 - \frac{\ln 2}{|\alpha|^2},
\end{align}
i.e., the deviation from 1 is suppressed for increasing $|\alpha|^2$.

\begin{figure}[t]
	\centering
	\includegraphics[width = 0.4\textwidth]{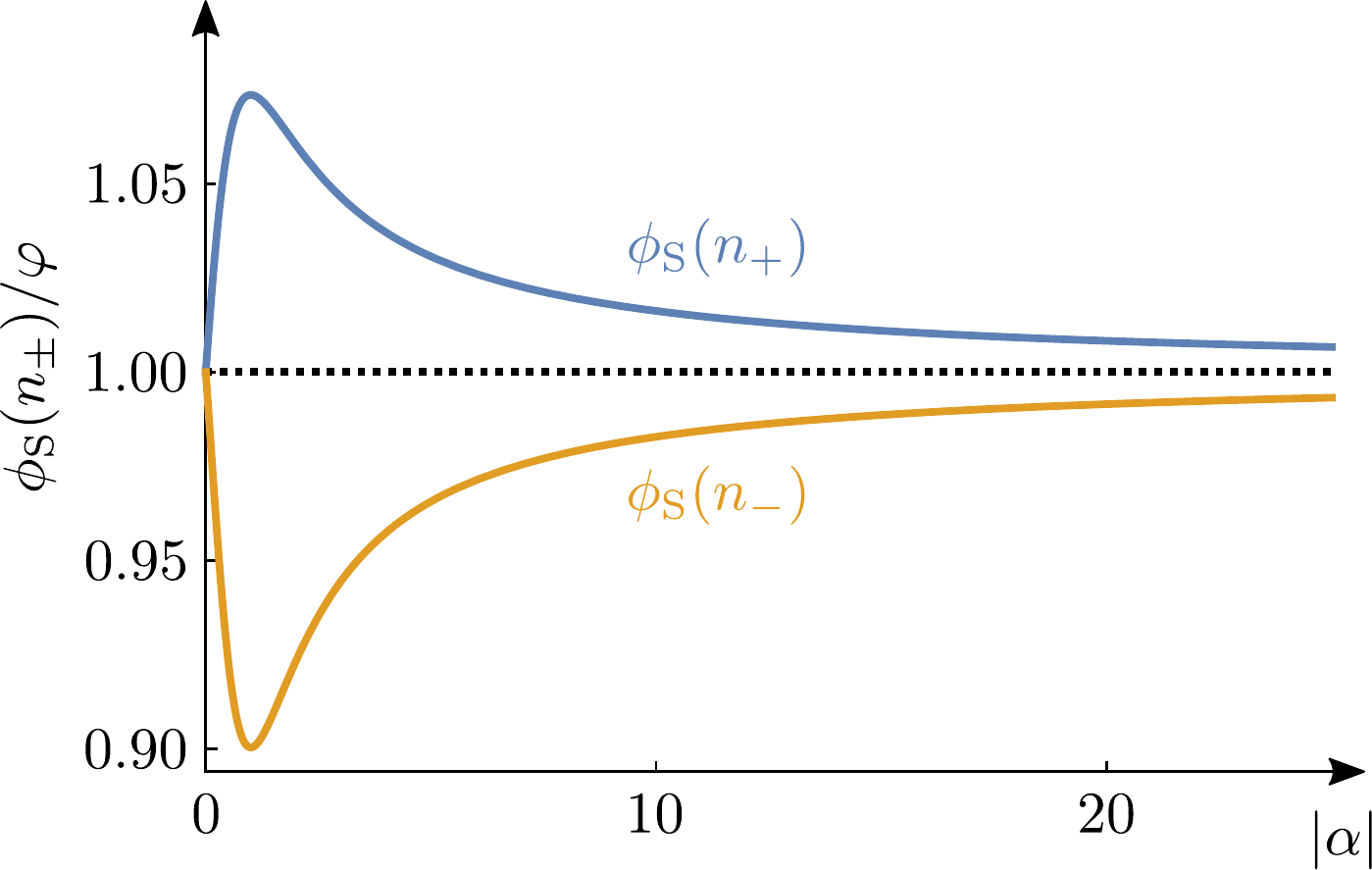}
	\caption{The deviation of the phase $\phi_{\rm S}(n)$ at $n = |\alpha|^2 \pm |\alpha|$ from the assumption $\phi_{\rm S}(n) = \varphi \approx -1.08$.
	The sweep speed is tuned such that $P_{\rm LZ}(|\alpha|^2) = \frac{1}{2}$ and thus $\phi_{\rm S}(|\alpha|^2) = \varphi$.}
	\label{fig:phis}
\end{figure}

Next we investigate in a similar way the $n$-dependence of $\phi_{\rm S}$ which we above also assumed to be constant, $\phi_{\rm S}(n) = \varphi \approx -1.08$.
Using the same sweep speed $v_f$ such that $P_{\rm LZ}(|\alpha|^2) = \frac{1}{2}$ we calculate the phases $\phi_{\rm S}(n_\pm)$ using Eq.~(\ref{stokesPhase}).
In Fig.~\ref{fig:phis} we show the result, where we normalized the two phases with $\varphi$.
We see again that the deviation from the ideal condition decreases monotonically for $|\alpha|^2 \gtrsim 1$.

Finally, we investigate the dynamical phases $\phi_d(n)$, for which we assumed $d(n) \equiv \phi_d(n+1) - \phi_d(n) = \pi$.
Assuming that the sweep is performed using the detuning extrema $-\hbar \omega \pm \hbar k$ (with $k < \omega$, see Fig.~\ref{fig:spectrum}) we find
\begin{align}\label{phaseDiff}
	d(n) \approx \frac{2\hbar A^2}{v^*} \left[\ln \left( \frac{k ^2}{A^2[n+1]}\right)  - \frac{1}{2[n+1]}\right],
\end{align}
valid in the limit $A^2n \ll k^2$ and at $n \gtrsim 1$, where $v^* = v_fv_s/(v_f + v_s)$ is an average of the two different sweep speeds, with $v_s$ the ``slow'' adiabatic sweep speed.
The derivative with respect to $n$ of this phase difference,
\begin{equation}
d'(n) \approx  \frac{2\hbar A^2}{ v^*}\left(- \frac{1}{n+1} + \frac{1}{2[n+1]^2}\right),
\end{equation}
decreases faster than $1/\sqrt n$ for increasing $n$, but the total error in $d(n)$ at $n_\pm = |\alpha|^2\pm |\alpha|$ is a cumulative error, contributed to by all $n$ between $|\alpha|^2$ and $n_\pm$.
To arrive at an estimate for the typical error in $d(n)$ we thus need to sum over all contributing phase differences from $|\alpha|^2$ to $n_\pm$, yielding approximately
\begin{align}
{} & {} \pm \frac{2\hbar A^2}{ v^*} \left(-\frac{1}{|\alpha|^2+1} + \frac{1}{2[|\alpha|^2+1]^2}\right)\sum_{l=1}^{|\alpha|}l \nonumber\\
= {} & {} \pm \frac{\hbar A^2}{ v^*} \left(-\frac{1}{|\alpha|^2+1} + \frac{1}{2[|\alpha|^2+1]^2}\right)\left(|\alpha|^2 + |\alpha|\right). \label{eq:phaseError}
\end{align}
We see that this contribution to the infidelity of our protocol does not decrease as a function of increasing $|\alpha|^2$: for large $|\alpha|$ it approaches $\hbar A^2/v^*$.
In principle one could tune $\hbar A^2/v^*$ as small as desired, but, due to the conditions (i) $d(|\alpha|^2) = \pi$ and (ii) $\hbar A^2|\alpha|^2/v_s \gg 1$, that would require an exponential increase of $k/A|\alpha|$, see Eq.~(\ref{phaseDiff}), and a very large ratio $\hbar k^2/v_s$.
These requirements are not in contradiction with any of our other assumptions, but might be inconvenient from a practical point of view.

We thus showed how our proposed protocol conceptually works, and we demonstrated that the most important intrinsic inaccuracies can in principle be tuned to be of arbitrarily small importance, e.g., by going to large $|\alpha|^2$ and small $\hbar A^2/v^*$.
However, deviations from the other assumptions we made, such as the validity of the adiabatic-impulse model, will also contribute to the infidelity of the protocol.
In order to investigate their importance quantitatively, we will present numerical simulations of the protocol in Sec.~\ref{sub:numerics}, showing that fidelities of $\sim 0.99$ are theoretically indeed achievable.

So far, we used the qubit mainly as a tool to tune the spectrum and sweep the system through multiple level crossings.
As an example, we showed how initialization of the system in $\ket{0}\otimes\ket{\alpha}$ can produce an odd cat state of the photon field after projecting the final qubit state to $\ket{0}$.
We could, however, also assume a more general initial state of the qubit, $\ket{\chi}_{\rm qub} = a\ket{0} + b\ket{1}$, and investigate to what extent and in what way the quantum information in this initial qubit state is transferred to the photon field during the sweep protocol.

It is straightforward to focus on the same pair of levels $\{ \ket{0,n}, \ket{1,n+1} \}$ as before and use Eq.~(\ref{eq:finalstate}) to write down the final state after initializing in $\ket{1,n+1}$, assuming the same ideal conditions as in (\ref{eq:final0n}),
\begin{align}
	\ket{1,n+1} \to -e^{-in\omega T}\Big [ {} & {} \sin\left(\frac{n}{2}\pi\right) \ket{1,n+1} \nonumber\\
	{} & {}  + ie^{i\varphi}\cos\left(\frac{n}{2}\pi\right)\ket{0,n} \Big].\label{eq:final1n+1}
\end{align}
We see that now for even $n$ the population is fully transferred to $\ket{0,n}$ and for odd $n$ to $\ket{1,n+1}$.
This means that initialization in $\ket{1}\otimes\ket{\alpha}$ would produce a final photon state with a perfect \emph{even} parity, whereas initialization in $\ket{0}\otimes\ket{\alpha}$ yielded a perfect \emph{odd} state.
Initializing in $\ket{\chi}_{\rm qub}\otimes\ket{\alpha}$ will thus yield a photonic state that is in a superposition of a perfectly odd state (with an amplitude proportional to $a$) and an even state (with amplitude proportional to  $b$).
Since the even and odd cat states $\ket{\Psi_\pm(\alpha)}$ form a good orthogonal basis for cat-based quantum information applications~\cite{Ralph2003,Lund2008}, where $\ket{\chi}_{\rm cat} = a\ket{\Psi_-(\alpha)} + b\ket{\Psi_+(\alpha)}$ encodes the same quantum information as $\ket{\chi}_{\rm qub}$, this suggests that our protocol might provide a way to transfer the quantum information coherently from the qubit to the photon field.

We thus initialize in $\ket{\chi}_{\rm qub}\otimes\ket{\alpha}$ and perform the same sweep protocol as before, assuming the same hypothetical ideal conditions.
If we again selectively measure the final state of the qubit, only accepting the outcome $\ket{0}$, we will project the photon field to the state
\begin{align}
	- {} & {} ia\ket{\Psi_-(\beta)} \nonumber\\
	- {} & {} ib e^{i\varphi} \frac{e^{-\frac{1}{2}|\beta |^{2}}}{\cal N}  \sum _{n=0}^{\infty}\frac{\alpha}{\sqrt{n+1}}\left[ \frac{\beta^{n}}{\sqrt {n!}} +  \frac{(-\beta)^{n}}{\sqrt {n!}} \right] \ket{n}.\label{eq:finaltotal}
\end{align}
We see that the even part of the field is nearly a cat state, the weight of each $n$-photon component being slightly off since it originated from the component $\ket{n+1}$ in the initial coherent state $\ket{\alpha}$ of the field.
The modulo square of the overlap of the part of the photon field proportional to $b$ with the state $-ie^{i\varphi}\ket{\Psi_+(\beta)}$ can be evaluated numerically: we find that it is $0.99$ for $|\alpha| \approx 5$ and approaches 1 monotonically for increasing $|\alpha|$.
Therefore we conclude that with high fidelity the final photon state approaches
\begin{align}
	-i \big[ a \ket{\Psi_-(\beta)} + e^{i\varphi} b \ket{\Psi_+(\beta)} \big].
\end{align}
The phase $\varphi \approx -1.08$ is known and can thus be compensated for, meaning that our sweep protocol indeed provides a means to coherently transfer quantum information from an actual two-level system to a superposition of even and odd cat states in a photon field.
Of course, the final state (\ref{eq:finaltotal}) was derived under the same assumptions concerning $P_{\rm LZ}(n)$, $\phi_{\rm S}(n)$, and $\phi_d(n)$ as (\ref{eq:final0n}), i.e., the deviations from these ideal conditions investigated before will affect the fidelity of this transfer protocol in a way that should be quantitatively similar.

\section{\label{sub:numerics}Numerical simulations}

We solved the time-dependent Schr\"{o}dinger equation using the function \texttt{sesolve} from the ``Quantum Toolbox in Python'' (QuTiP) package~\cite{Johansson2011,Johansson2013}.
We initialize the system in a direct product state of one of the two qubit basis states and a coherent state of the photon field $\ket{\alpha}$, using the basis states $\ket{0,n}$ and $\ket{1,n}$, where we cut off the Hilbert space for $n \geq 2*|\alpha|^2 + 10$, and work in units where $\hbar = 1$.
Then we evolve the system using the Hamiltonian as given in (\ref{transHamiltonian}) with
\begin{align}
	\Delta(t) = 
	\begin{cases}
		-\omega -k+v_f t, & \text{for } 0 \leq t \leq \tau_1, \\
		-\omega+k-v_s(t-\tau_1), & \text{for } \tau_1 \leq t \leq \tau_2, \\
		-\omega-k+v_f(t-\tau_2), & \text{for } \tau_2 \leq t \leq \tau_3,
	\end{cases}\label{eq:protocol}
\end{align}
with $\tau_1 = 2k/v_f$, $\tau_2 = \tau_1 + 2k/v_s$, and $\tau_3 = \tau_2 + 2k/v_f$.
This results in the sweep pattern as shown in Fig.~\ref{fig:spectrum}(b), with the detuning extrema $-\omega \pm k$.

We will first present numerical results for $|\alpha|^2 = 10$.
We set $A = 0.003$ (in units of $\omega$), which will allow for large $k/A|\alpha|$ while still satisfying the condition $k \leq \omega$.
Then we set the other simulation parameters as follows: $v_f$ is calculated using Eq.~\eqref{eq:PLZCond}, demanding that $P_{\rm LZ}(10) = \frac{1}{2}$, and $v_s$ and $k$ are tuned together to satisfy $d(10) = \pi$, see Eq.~\eqref{phaseDiff}, and $\phi_d(10) = -2\phi_{\rm S}(10)$ as closely as possible, which corresponds to fixing the ``offset'' phase $\phi_d(0) = -2\varphi$ in the ideal picture presented above.
This yielded $v_f \approx 8.9741 \times 10^{-4}$, $v_s \approx 4.6965 \times 10^{-5}$, and $k = 0.50017$.

\begin{figure}[t]
	\centering
	\includegraphics[width = 0.45\textwidth]{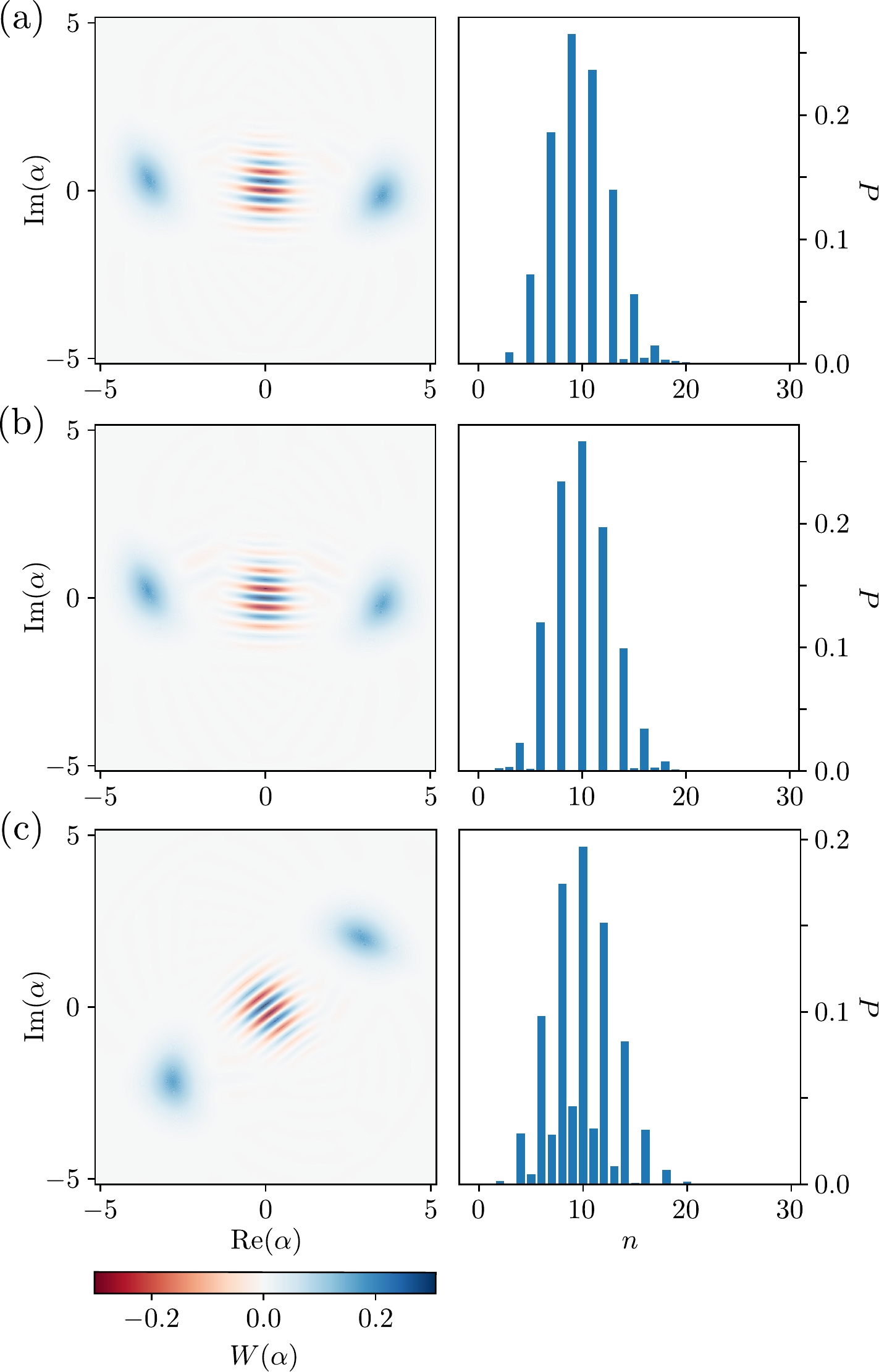}
	\caption{Numerically calculated final photon states after performing the sweep protocol (\ref{eq:protocol}) and projecting the qubit to the state $\ket{0}$.
		In the left panels we show their Wigner distribution function $W(\alpha)$ and in the right panels their photon number distribution function $P_n$.
		In all simulations we used $|\alpha|^2 = 10$ and $A = 0.003$.
		The sweep speeds $v_{s,f}$ were chosen such that $P_{\rm LZ}(10) = \frac{1}{2}$ and $d(10) = \pi$.
		(a,b) Fine-tuning the parameter $k$ to $k = 0.50017$ yields (a) an odd cat state with fidelity 0.986 if the qubit was initialized in $\ket{0}$ and (b) an even cat state with fidelity 0.989 if it was in $\ket{1}$.
		(c) Not fine-tuning $k$ yields a cat state of a more general form, see Eq.~(\ref{generalCatState}); in this case $k=0.49$ resulted in $\theta = 0.713\,\pi$ with fidelity 0.988.}
	\label{fig:oddEven}
\end{figure}
In Fig.~\ref{fig:oddEven} we show final photon states after the sweep protocol, assuming the qubit to have been projected to $\ket{0}$.
The left panels show the Wigner distribution function $W(\alpha)$ of the state, all of them showing the characteristics of a cat state, including the typical fringes around $\alpha = 0$ signaling quantum-mechanical interference between the state's two main components.
The right panels show the photon number distribution $P_n$ of the final state.

Fig.~\ref{fig:oddEven}(a) presents the final photon state that results after initializing the qubit in $\ket{0}$.
In the right panel we see that state has almost perfect odd parity, suggesting that it is indeed close to an odd cat state, as predicted above.
We can calculate numerically the overlap of this final state with the desired odd cat state, as a measure of the fidelity of our protocol, yielding $F_- = |\braket{\Psi_f|\Psi_-(\beta)}|^2 \approx 0.986$, where $|\beta| = \sqrt{10}$.
In Fig.~\ref{fig:oddEven}(b) we show the resulting final photon state after the same sweep procedure, but now having initialized the qubit in $\ket{1}$.
The parity of this state is indeed almost exactly opposite, and as fidelity of the thusly created even cat state we find $F_+ \approx 0.989$.
As expected, the Wigner distribution functions of the two states shown in Fig.~\ref{fig:oddEven}(a,b) look very similar. The main difference is in the phase of the interference fringes appearing around $\alpha = 0$, which is set by the relative sign of the two components constituting the cat state.
Finally, in Fig.~\ref{fig:oddEven}(c) we show the resulting state if the parameter $k$ is not fine-tuned but simply fixed to some value, we set $k = 0.49$, after which only $v_f \approx 8.9741 \times 10^{-4}$ and $v_s \approx 4.6705$ are adjusted to satisfy $P_{\rm LZ}(10)=\frac{1}{2}$ and $d(10) = \pi$, following Eqs.~(\ref{eq:PLZCond}) and \eqref{phaseDiff}.
This means we no longer ensure that $\phi_d(10) = -2\phi_{\rm S}(10)$, which introduces an extra phase shift in the $n$-dependent oscillations of all final amplitudes, see Eq.~(\ref{eq:final0nfull}).
This phase shift results in a final state that is still close to a cat state, but now of the more general form
\begin{equation}\label{generalCatState}
	\ket{\Psi_\theta(\beta)} = \frac{1}{\cal N} \left( \ket{\beta} + e^{i\theta} \ket{-\beta} \right),
\end{equation}
where $\theta$ is no longer necessarily $\pi$ or 0.
For our particular choice of $k$ we found a maximum overlap with the cat state with $\theta = 0.713\,\pi$, yielding a fidelity of $F_\theta \approx 0.988$.

\begin{figure}[t]
	\centering
	\includegraphics[width=0.47\textwidth]{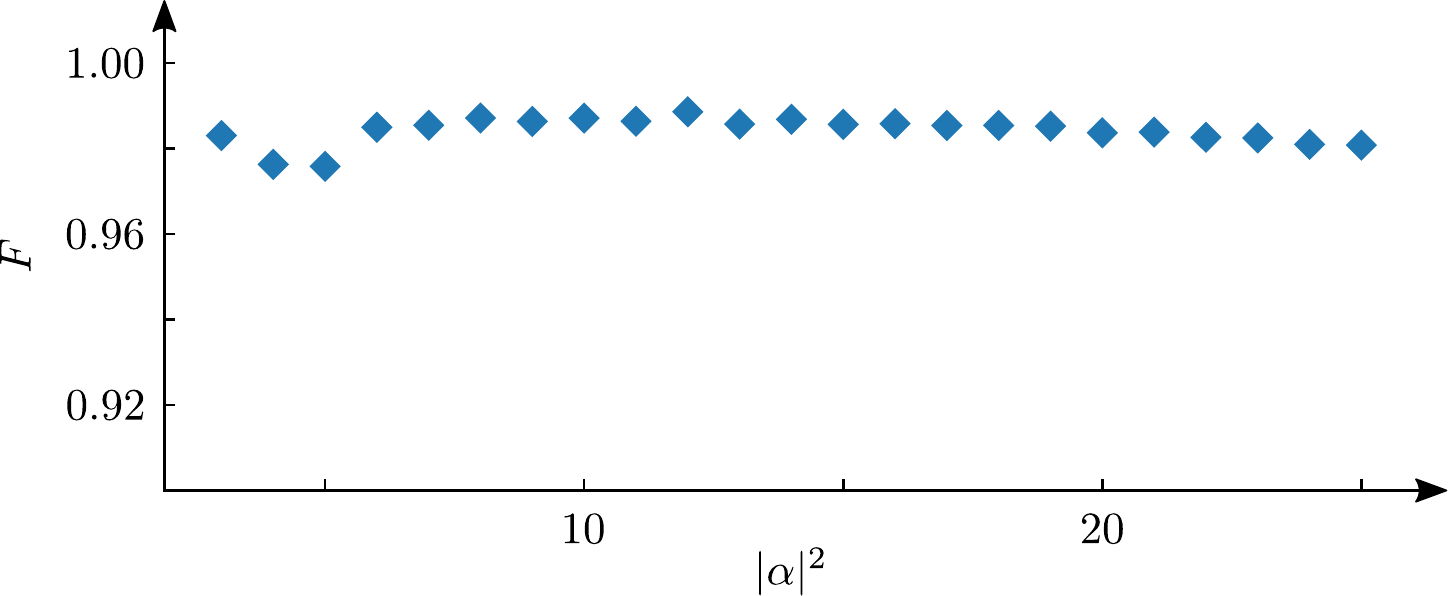}
	\caption{Fidelity of the resulting final cat state as a function of $|\alpha|^2$ using the same parameters $A=0.003$ and $k=0.50017$ throughout, while adjusting  $v_f$ and $v_s$ such that the conditions $P_{\rm LZ}(|\alpha|^2)=\frac{1}{2}$ and $d(|\alpha|^2) = \pi$ remain fulfilled.}
	\label{fig:fidnfunc}
\end{figure}

We thus see that our protocol is in principle indeed able to produce Schr\"{o}dinger cat states with fidelities up to $\sim 0.99$, already at moderate $|\alpha|$.
Next, we investigate how the fidelity of the protocol depends on $|\alpha|$ by varying $|\alpha|^2$ from 3 to 25.
We initialize the qubit in $\ket 0$ and perform the sweep procedure using the same parameters $A = 0.003$ and $k = 0.50017$ as before, while adjusting $v_f$ and $v_s$ such that $P_{\rm LZ}(|\alpha|^2)=\frac{1}{2}$ and $d(|\alpha|^2) = \pi$ for each $|\alpha|^2$.
After projecting the qubit to $\ket 0$ we find the cat state $\ket{\Psi_\theta(\beta)}$ that has largest overlap with the final photon state and calculate the modulo square of that overlap to determine the fidelity $F$.
In Fig.~\ref{fig:fidnfunc} we plot $F$ as a function of $|\alpha|^2$, and we see that the fidelity is around 0.98 for all $|\alpha|^2$ in the range plotted, with a slight decrease for larger $|\alpha|^2$. This is a sign that the assumption of $A|\alpha|/k \ll 1$, which is one of the conditions for the adiabatic-impulse model to be a good approximation, is starting to become questionable. 

\section{\label{sec:realization}Discussion}

We demonstrated that our protocol can generate cat states in a photon cavity with fidelities up to $\sim 0.99$ and could possibly also be used to coherently transfer quantum information between a qubit and a superposition of different, orthogonal cat states in the cavity.
However, so far we assumed both the cavity and the qubit to be lossless and coherent at all times.
In reality, both subsystems are coupled to many environmental degrees of freedom, leading to photon loss from the cavity, typically characterized with the decay rate $\kappa$, and qubit decoherence characterized with the rate $\gamma$.
Roughly speaking, good coherence during the full protocol will thus be guaranteed if the total time of the procedure $T \sim 2\hbar k/v_s \ll \kappa^{-1}, \gamma^{-1}$, where $v_s$ should be small enough to ensure that the second crossing is adiabatic.
It is therefore straightforward to investigate the typical coherence properties of a few obvious candidate systems for implementing our protocol and compare them in the context of this requirement.

The paradigmatic system providing a well-controllable qubit coupled to a single mode of the photon field is the circuit-QED setup of a superconducting qubit coupled to a transmission line resonator~\cite{Blais,Wallraff2004}.
In such hybrid systems coupling strengths of $A/2\pi \sim 100~$MHz can easily be reached~\cite{Blais2020}, which, assuming $|\alpha|^2 \sim 10$, leads to the constraint $v_s \lesssim 25~\mu$eV/ns.
Typical resonator frequencies are of the order $\omega/2\pi \sim 10$~GHz and picking $2k \approx \omega$, as we did in our numerical examples, then yields a lower limit $T \gtrsim 2$~ns, which is well below typical decay and dephasing times $\kappa^{-1}, \gamma^{-1} \sim 0.1$--$1~\mu$s.
In fact, depending on the choice of qubit (transmon, flux, phase, or charge), having such a strong coupling parameter $A$ might make it challenging to implement a sweep speed $v_f$ high enough to yield a Landau-Zener probability of $\frac{1}{2}$.
But since the minimum and maximum $T$ estimated above are still several orders of magnitude apart, there is enough room to work with significantly smaller coupling $A$.

Another, more recently developed class of hybrid systems that could be used to implement our idea is that of gate-defined semiconductor quantum dots coupled to a superconducting cavity~\cite{Burkard2020}.
This would allow to perform the protocol using a spin-based qubit instead of a superconducting qubit, which provides potentially superior qubit coherence properties.
Direct spin-photon coupling is weak, typically on the peV scale, but the effective coupling strength can be significantly enhanced by using multi-electron spin qubits instead, where the basis states are spin-charge mixtures that couple much more efficiently to the photon field~\cite{Burkard2020}.
The most advanced example in this field is the triple-dot three-electron exchange-only qubit~\cite{DiVincenzo2000, Taylor2013, Russ2017}, which provides fast all-electric control and potentially strong coupling to the cavity mode~\cite{Srinivasa2016a}.
Recently the coherent coupling between such a qubit and a microwave cavity has been demonstrated experimentally~\cite{Landig2018a}, yielding the device parameters $\omega/2\pi \approx 4.5$~GHz, $\kappa^{-1}\approx 20$~ns, and $\gamma^{-1} \approx 0.1~\mu$s, while providing coupling strengths up to $A/2\pi \approx 31~$MHz.
The same rough estimate as above then yields the constraint $v_s \lesssim 2~\mu$eV/ns and thus a lower limit $T \gtrsim 9$~ns, which is within the reported decoherence times.

\section{Conclusion}\label{sec:conc}

We presented a conceptually new way to create Schr\"{o}dinger cat states in a photonic cavity using Landau-Zener-St\"{u}ckelberg interferometry, by coupling the cavity to a qubit and manipulating the qubit splitting as a function of time.
We show how our protocol can create cat states with a fidelity up to $\sim 0.99$ for $|\alpha|^2 \sim 10$, and how it could also be used to coherently transfer quantum information between the qubit and the photon field, where it can be stored in the form of coherent superpositions of orthogonal cat states.
We corroborated our presentation of the protocol with numerical simulations and finally discussed a few candidate hybrid systems that could be used to implement our idea.

\section*{Acknowledgments}

We gratefully acknowledge financial support from NTNU's Onsager Fellowship Program, as well as partial support through the Centers of Excellence funding scheme of the Research Council of Norway, project number 262633, QuSpin.
We thank A.~Kamra for giving useful feedback on our manuscript in preparation.

\end{document}